\documentclass[5p,times]{elsarticle}   % [review]

\usepackage{lipsum}

\usepackage{caption}

\usepackage{listings}
 
\usepackage{subcaption} 

\usepackage{amsmath,graphicx}

\usepackage[dvipsnames]{xcolor}

\usepackage{array}
\usepackage[justification=centering]{caption}
\usepackage[font=normalsize]{caption}
\usepackage{listings}
\usepackage{epsfig}

\usepackage{hyperref}
\usepackage{breakurl}

\usepackage{lineno,hyperref}
\modulolinenumbers[5]

\newlength{\fslength}

\makeatletter
\newcommand{\funnyP}{%
    \setlength{\fslength}{\f@size pt}%
    \reflectbox{${\mbox{P}}$}\hspace*{-.300\fslength}\mbox{${\mbox{P}}$}%
}
\newcommand{\funnyPbar}{%
    \setlength{\fslength}{\f@size pt}%
    \reflectbox{$\bar{\mbox{P}}$}\hspace*{-.300\fslength}\mbox{$\bar{\mbox{P}}$}%
}
\newcommand{\funnyPbarM}{%
    \setlength{\fslength}{\f@size pt}%
    \reflectbox{$\bar{\mbox{P}}$}\hspace*{-.300\fslength}\mbox{$\bar{\mbox{P}}_{M}$}%
}
\makeatother

\usepackage{soul}

\begin{document}

\begin{frontmatter}

%\title{On the Productive Performance Portability of OpenACC}
\title{Portability Efficiency Approach for Calculating Performance Portability}

\author{Ami~Marowka}
\address{Parallel Research Lab\\
8 Rosh Pina, Petach Tikva, Israel. 49729 \\
amimar2@yahoo.com}

\begin{abstract}
The emergence of heterogeneity in high-performance computing, which harnesses under one integrated system several platforms of different architectures, also led to the development of innovative cross-platform programming models. Along with the expectation that these models will yield computationally intensive performance, there is demand for them to provide a reasonable degree of performance portability. Therefore, new tools and metrics are being developed to measure and calculate the level of performance portability of applications and programming models. 

\smallskip
The ultimate measure of performance portability is performance efficiency. Performance efficiency refers to the achieved performance as a fraction of some peak theoretical or practical baseline performance. {\it Application efficiency} approaches are the most popular and attractive performance efficiency measures among researchers because they are simple to measure and calculate. Unfortunately, the way they are used yields results that do not make sense, while violating one of the basic criteria that defines and characterizes the performance portability metrics. 

\smallskip
In this paper, we demonstrate how researchers currently use application efficiency to calculate the performance portability of applications and explain why this method deviates from its original definition. Then, we show why the obtained results do not make sense and propose practical solutions that satisfy the definition and criteria of performance portability metrics.
Finally, we present a new performance efficiency approach called {\it portability efficiency}, which is immune to the shortcomings of application efficiency and substantially improves the aspect of portability when calculating {\it performance portability}.  
\end{abstract}


\end{frontmatter}

%\linenumbers

\section{Introduction}

The increased use of contemporary heterogeneous systems continues to challenge the designers of modern cross-platform programming models. The main difficulty designers face is to achieve the three pillars of high-performance computing: performance, portability, and productivity which are in tension with each other \cite{Anzt, Gropp, DOE}.

To assess an application's performance portability degree, it must be measured and calculated empirically on a sufficient number of different platforms. Conducting experiments on ten platforms is certainly sufficient, while measurements carried out on three platforms will yield a very deficient assessment.

There are several preliminary steps before we calculate the performance portability of an application. First, it is important that we clarify to ourselves what the definition of the term {\it performance portability} really means. After that, determining which metric is most appropriate for our needs is required. Then, we need to choose a set of platforms of interest, and finally we must to choose the performance efficiency for the primary measurements.

Research engaged in finding new and better methods for examining and measuring performance portability is still ongoing. However, it seems that regarding the definition of the term performance portability, there is broad consensus \cite{Pennycook2}.

\begin{quote}
{\bf Definition: performance portability}

{\it  A measurement of an application's performance efficiency for a given problem that can be executed correctly on all platforms in a given set.}
\end{quote}

The definition explicitly states that performance efficiency is the ultimate measure of performance portability. Therefore, several approaches were proposed to measure performance efficiency alongside several metrics to calculate performance portability \cite{Marowka2}, \cite{Marowka3}, \cite{Marowka4}.  
The performance efficiency of a given application on a platform of interest is defined as follows:

\begin{quote}
{\bf Definition: Performance Efficiency}

{\it A measurement of an application's achieved performance as a fraction of a baseline performance.}
\end{quote}

When performance is usually measured by runtime or throughput, the baseline performance can be either the theoretical or practical peak performance, such as the theoretical peak throughput of a specific GPU or its roofline peak throughput  \cite{Bertoni}.  

Currently, two metrics are used to calculate the performance portability of an application and two types of performance efficiencies, {\it architectural efficiency} and {\it application efficiency}, which use different performance baselines to calculate performance efficiency \cite{Marowka4}. Baseline performances are mainly divided into two categories: theoretical and practical. For example, the two common architectural efficiency baselines are the theoretical peak performance of the platform of interest and the practical roofline peak performance of the platform of interest.
Application efficiency is a popular measure because it is simple and easy to use \cite{Daniel}-\cite{Deakin3}. All that is required is to measure the achieved runtime of the application on the given platform, and then to calculate its fraction relative to the runtime of the fastest known implementation of the application on the same platform.

The problem is that we can never be sure whether we have the fastest implementation at hand. And so it can happen that immediately after we have published our research, a faster implementation is found which makes the results of our findings obsolete.

Furthermore, in all the recent studies of performance portability of applications that are based on application efficiency approach, 
%according to recent studies on application efficiency, 
researchers always chose as the baseline performance the performance of the implementation that showed the best performance from three or four implementations studied in their {\it current} research and not from those known in the literature \cite{Daniel}-\cite{Deakin3}.
If we add the observation that different studies use different compilers, compiler options, and input sizes-and that the source codes are not always available-it is clear that this situation leads to non-uniformity and incoherence of the results, and difficulties in reproducing them.

In this paper we concentrate on how application efficiency has been used since it was first proposed  \cite{Pennycook2}. We present in detail, with the help of demonstrations, how it has been used, which will clarify the deficiencies inherent in the current calculation method and their consequences. After that, we present a few solutions that do not violate the definition and criteria of performance portability metrics.

Finally, we present a new performance efficiency approach called {\it portability efficiency}, which is immune to the shortcomings of application efficiency and better reflects the aspect of {\it performance portability}.

In addressing these goals, we make the following contributions:

\begin{itemize}
%\item We detail the criteria that a performance portability metric should satisfy.
%\item We detail the formal definition of the {\funnyPbar} metric and definitions of the architectural and application  efficiencies.
\item We demonstrate how application efficiency has been used and the deficiencies arising from this method of calculation.
\item We present flexible solutions that are not affected by the deficiencies found in the current calculation method.
\item We present a new performance efficiency approach called {\it portability efficiency} which is immune to the shortcomings of application efficiency.
\end{itemize}

We use the {\funnyPbar} metric for calculating performance portability \cite{Marowka3, Marowka4}  in our demonstrations, which is based on the arithmetic mean, simply because it is more mathematically and practically correct. However, all the problems of measuring and calculating application efficiency which are presented in this paper, and their solutions, are also correct for every performance portability metric that has been proposed so far in the scientific literature.

The rest of the paper is structured as follows.  Section 2 reviews the criteria of the {\funnyPbar} metric, its definition, and the definitions of the architectural and application efficiencies. Section 3 presents related works.  Section 4 demonstrates the current method of calculating performance portability using application efficiency and its shortcomings. Section 5 describes appropriate solutions for calculating performance portability using application efficiency. Section 6 presents an undesirable solution which was used in a recent study.
Section 7 presents a new performance efficiency approach called portability efficiency, and Section 8 presents the conclusions.

\section {The {\funnyPbar} Metric}

This section presents the criteria and definition of the {\funnyPbar} metric and the definitions of architectural and application efficiencies.

Given a set of supported platforms $S \subseteq H$, the set of criteria of the {\funnyPbar} metric defines it to be:

\begin{enumerate}
\item{measured specific to a set of platforms of interest $S$}
\item{independent of the absolute performance across $S$}
\item{zero if none of the platforms is supported.}
\item{increasing or decreasing if performance increases 
or decreases on any platform in $S$}
\item{directly proportional to the sum of scores across $S$}
\end{enumerate}

The {\funnyPbar} metric is defined as the arithmetic mean of an application's performance efficiency observed across a set of platforms from the same architecture class. Formally, for a given supported set of platforms $S \subseteq H$ from the same architecture class, the performance portability of a case-study application $a$ solving problem $p$ is: 

\begin{align}
{\funnyPbar}(a, p, S, H) = 
\begin{cases}
  \frac{\sum_{i \in S}^{} {e_{i}(a, p)}}{|S|}     & \text{if $|S| > 0$} \\ 
  \text{$0$}                                      & \text{otherwise} 
\end{cases}
\end{align}

where $S:= \{i \in H|e_{i}(a, p) > 0\}$ and 
$e_{i}(a, p)$ is the performance efficiency of case-study application $a$ solving problem $p$ on platform $i$.

Two performance efficiency approaches have been proposed to date in the scientific literature: {\it application efficiency} and {\it architectural efficiency}. These two approaches present two different perspectives on the relative performance of a given application running on a particular platform and both yield different scores. Each of them examines the performance of a given application in relation to different reference performances. Application efficiency is measured in relation to the performance of the fastest known implementation on that platform, while architectural performance is measured in relation to the theoretical or practical performance that can possibly be achieved on the given platform. 

\smallskip
Now let us define these two approaches formally.

\begin{quote}
{\bf Definition: architectural efficiency}

{\it The applications achieved throughput on a given platform normalized relative to the peak throughput of the given platform}.
\end{quote}

\begin{quote}
{\bf Definition: application efficiency}

{\it The achieved performance, on a given platform, normalized relative to the {\textbf {best-known performance of an application's implementation on the same platform}.}}
\end{quote}

Here we call the reader's attention to the fact that, in the definition of the application efficiency, the phrase ``{\it the best-known performance of an application's implementation on the same platform''} is highlighted for good reason. As we see in the next section, studies that used application efficiency to calculate performance portability up to the present day did not bother to find the best-known performance of the application's implementation on the same platform. Instead they chose the best-known performance of the application's implementation on the same platform {\it in the current study}, which almost never represents the existing best-known performance of the application's implementation on the same platform. This is the mistake that leads to calculations that do not make sense, as demonstrated and explained in the next section.

\section {Related Works}

In this section we briefly describe a sample of studies that used application efficiency.

In \cite{Daniel}, Daniel and Panetta proposed a metric called Performance Portability Divergence (PD) to quantify the performance portability of an application across multiple machine architectures. The authors showed that the metric developed by Pennycook et al. \cite{Pennycook2}, {\funnyP}, is sensitive to problem size and therefore proposed a new metric to address this deficiency. The definition of the PD metric is based on the definition of code divergence D(A), which is the average of the pairwise distances between applications in the set of codes A as proposed by Harrell et al. \cite{Harrell}. Mathematically speaking, PD is the complement of {\funnyP} when the performance efficiency is replaced by the average of the differences of the complement of performance efficiency for different input sizes. 

The authors demonstrated the use of their metric by experiments on two CPUs (Xeon E5-2630 v4 and Xeon E5-2699 v4) and two GPUs (NVIDIA Tesla K80 and NVIDIA Tesla P100). They used eight scientific codes implemented using the Kokkos and OpenACC parallel programming models to calculate the performance portability of these models across the CPUs and GPUs used in their experiments. The performance portability results obtained using the PD metric were analyzed and compared with the performance portability obtained by the {\funnyP} metric. However, the calculations of these two performance portability metrics are based on the principles of application efficiency approach thus suffer from the same problems described in Section 4.  

In \cite{Harrell}, Harrell et al. proposed a new definition of productivity and an associated metric that captured the development efforts to optimize and port an application across different platforms. The metric was called code divergence, D(A), which is the average of the pairwise distances between all the applications in a set A, where distance is the change in the number of source lines of code normalized to the size of the smaller application. The authors used their metric and the {\funnyP} metric, using the application efficiency approach, to study the three Ps (Performance, Portability, and Productivity) of three scientific applications when ported and optimized for several performance portability frameworks. They used a logging tool to collect data on the development process and its associated methodology. Despite the enormous effort invested in this study, the authors could not draw any conclusions about the productivity of the tested performance portability frameworks.

\smallskip
In \cite{Deakin2}, Deakin et al. presented an extensive study of the performance portability of five mini-applications implemented using five parallel programming models across six CPUs, five GPUs, and one vector-based architecture. The calculation of performance portability in this study was carried out using the application efficiency approach, with the best performance efficiency of a non-portable programming model (CUDA or OpenCL) functioning as a performance efficiency baseline. The authors' intention to conduct extensive research was largely unrealized because of problems such as immaturity of the tested programming models or imperfection of compilers and runtime systems. Particularly noticeable was the omission of many implementations in CUDA and OpenCL, whose performance could have provided baseline performance efficiencies. 

In such cases, the authors chose the performance of the high-level abstraction model that exhibited the best performance, such as OpenMP or Kokkos, as the baseline performance, which necessarily produced biased results. In cases where an application has only a single implementation, then there is no other choice but to determine that {\funnyP} = 100\%, or to determine that the baseline performance will be the best performance of the application among the tested architectures H, rather than the best-known implementation. Five tables (Figures 1, 3, 5, 7 and 9) in \cite{Deakin2}
show the performance efficiency scores of each of the five applications in the study. It is apparent that on average, 25\% of the cells in each table are empty, which means that 25\% of the implementations are missing. Therefore, it was inevitable that the results indicated considerable distortion and an inability to estimate properly the performance portability of the applications being tested.

\smallskip
In \cite{Pennycook5}, Pennycook et al. using a molecular dynamics benchmark called miniMD from the Mantevo suite to develop implementations of OpenMP 4.5 for CPU (Intel Xeon Gold 6148) and GPU (NVIDIA P100) and to calculate their application efficiencies and their performance portability, {\funnyP}. For the CPU, the baseline performance used was the performance of an implementation of miniMD, called mxhMD, which was developed by the authors in a previous work. For the GPU, the baseline performance used was the performance of the Kokkos implementation of miniMD because no CUDA version of miniMD exists. Therefore, the application efficiencies of these two implementations were set to 100%.

\smallskip 
The implications of these constraints are as follows. In the absence of a version of CUDA, it is necessarily to set the application efficiency of the original and mxhMD versions to 0\%, and therefore their {\funnyP} scores are 0\%. These results cannot reflect anything significant about the performance portability of the original and mxhMD versions except that they do not yet have an implementation in CUDA. Furthermore, the application efficiencies of OpenMP were found to be 35.49\% and 100\% on the CPU and GPU respectively, which together yield a {\funnyP} score of 52.39\%. Because there is a three-fold gap between the CPU and GPU application efficiencies, this is inevitably a biased result.

\smallskip
In \cite{Kirk}, Kirk et al. studied the performance portability of TeaLeaf, a mini-application from the Mantevo suite that solves the linear heat-conduction equation. Two performance portability framework implementations were used (Kokkos and RAJA) on a set of target architectures that included two CPUs (Xeon E5-2660 and KNL) and one GPU (P100). RAJA and Kokkos showed {\funnyP} scores of 77\% and 53\% respectively based on the application efficiency approach and scores of 61\% and 41\% respectively based on the architectural efficiency approach.

\smallskip
Siklosi et al. \cite {Siklosi} examined the performance of Stencil applications on hybrid CPU-GPU systems. They found that using the {\funnyP} metric to calculate the performance portability of applications is not intuitive. In their opinion, the reason for this is that if architecture efficiency is used, then the {\funnyP} metric tends to track the low values and therefore the improvement of a hybrid system is not reflected in the calculated {\funnyP} score. However, when using application efficiency, a hand-tuned baseline implementation is required, which to the best of their knowledge does not exist.

\section{Application Efficiency Calculation}

This section demonstrates the current method of performance portability calculation using application efficiency and its flaws.

\begin{table*}%[!htb]
%\begin{center}
\centering
\caption{CloverTree Mini-app Running Times (left) and Application Efficiencies \\ \& Performance Portability (right).}
\setlength\extrarowheight{2.0pt}
%\vspace{.5cm}
\hspace{-2.0cm}
\begin{minipage}{.27\linewidth}
%\caption{CloverTree Mini-app Running Times.}
%\centering
%\begin{center}
%\footnotesize
%\resizebox{0.7\columnwidth}{0.08\textheight}{% % !
%\resizebox{0.7\columnwidth}{!}{% 
%\setlength\extrarowheight{2.5pt}
\begin{tabular}{|l|l|l|l|}
\hline  
\multicolumn{1}{|c|}{} 
&\multicolumn{1}{|c|}{A100} 
&\multicolumn{1}{|c|}{P100} 
&\multicolumn{1}{|c|}{MI250} \\ \cline{1-4}
\hline  
\hline  
\multicolumn{1}{|l|}{OpenACC} 
&\multicolumn{1}{|c|}{30} 
&\multicolumn{1}{|c|}{50} 
&\multicolumn{1}{|c|}{60} \\ \cline{1-4}
\hline  
\hline  
\multicolumn{1}{|l|}{OpenMP} 
&\multicolumn{1}{|c|}{40} 
&\multicolumn{1}{|c|}{25} 
&\multicolumn{1}{|c|}{50} \\ \cline{1-4}
\hline  
\hline  
\multicolumn{1}{|l|}{Kokkos} 
&\multicolumn{1}{|c|}{60} 
&\multicolumn{1}{|c|}{75} 
&\multicolumn{1}{|c|}{40} \\ \cline{1-4}
\hline  
\end{tabular}
%}
%\end{center}
%\label{table:t1}
\end{minipage}
%\end{table}  
%\begin{table}
\hspace{0.7cm}
\begin{minipage}{.25\linewidth}
%\caption{CloverTree Mini-app Application Efficiencies \\ \& Performance Portability.}
%\begin{center}
%\footnotesize
%\resizebox{0.7\columnwidth}{0.08\textheight}{% % !
%\resizebox{0.7\columnwidth}{!}{% 
%\setlength\extrarowheight{2pt}
%\centering
\begin{tabular}{|l|l|l|l|l|}
\hline  
\multicolumn{1}{|c|}{} 
&\multicolumn{1}{|c|}{A100} 
&\multicolumn{1}{|c|}{P100} 
&\multicolumn{1}{|c|}{MI250} 
&\multicolumn{1}{|c|}{{\funnyPbar}} \\ \cline{1-5}
\hline  
\hline  
\multicolumn{1}{|l|}{OpenACC} 
&\multicolumn{1}{|c|}{100\%} 
&\multicolumn{1}{|c|}{50\%} 
&\multicolumn{1}{|c|}{66\%} 
&\multicolumn{1}{|c|}{72\%} \\ \cline{1-5}
\hline  
\hline  
\multicolumn{1}{|l|}{OpenMP} 
&\multicolumn{1}{|c|}{75\%} 
&\multicolumn{1}{|c|}{100\%} 
&\multicolumn{1}{|c|}{80\%} 
&\multicolumn{1}{|c|}{85\%} \\ \cline{1-5}
\hline  
\hline  
\multicolumn{1}{|l|}{Kokkos} 
&\multicolumn{1}{|c|}{50\%} 
&\multicolumn{1}{|c|}{33\%} 
&\multicolumn{1}{|c|}{100\%} 
&\multicolumn{1}{|c|}{61\%} \\ \cline{1-5}
\hline  
\end{tabular}
%}
%\end{center}
%\label{table:t2}
\end{minipage}
%\end{center}
\end{table*}  

\begin{table*}%[!htb]
\begin{center}
\caption{CloverTree Mini-app Running Times (left) and \\Application Efficiencies \& Performance Portability (right) including SYCL.}
%\vspace{.5cm}
\hspace{-0.8cm}
\setlength\extrarowheight{2.0pt}
\begin{minipage}{.27\linewidth}
%\caption{CloverTree Mini-app Running Times.}
%\begin{center}
%\footnotesize
%\resizebox{0.7\columnwidth}{0.08\textheight}{% % !
%\resizebox{0.7\columnwidth}{!}{% 
%\setlength\extrarowheight{2.0pt}
\begin{tabular}{|l|l|l|l|}
\hline  
\multicolumn{1}{|c|}{} 
&\multicolumn{1}{|c|}{A100} 
&\multicolumn{1}{|c|}{P100} 
&\multicolumn{1}{|c|}{MI250} \\ \cline{1-4}
\hline  
\hline  
\multicolumn{1}{|l|}{OpenACC} 
&\multicolumn{1}{|c|}{30} 
&\multicolumn{1}{|c|}{50} 
&\multicolumn{1}{|c|}{60} \\ \cline{1-4}
\hline  
\hline  
\multicolumn{1}{|l|}{OpenMP} 
&\multicolumn{1}{|c|}{40} 
&\multicolumn{1}{|c|}{25} 
&\multicolumn{1}{|c|}{50} \\ \cline{1-4}
\hline  
\hline  
\multicolumn{1}{|l|}{Kokkos} 
&\multicolumn{1}{|c|}{60} 
&\multicolumn{1}{|c|}{75} 
&\multicolumn{1}{|c|}{40} \\ \cline{1-4}
\hline  
\hline  
\multicolumn{1}{|l|}{SYCL} 
&\multicolumn{1}{|c|}{50} 
&\multicolumn{1}{|c|}{40} 
&\multicolumn{1}{|c|}{30} \\ \cline{1-4}
\hline  
\end{tabular}
%}
%\end{center}
\label{table:t1}
\end{minipage}
%\end{table}  
%\begin{table}
\hspace{0.7cm}
\begin{minipage}{.4\linewidth}
%\caption{CloverTree App. Efficiencies \& Performance Portability.}
%\begin{center}
%\footnotesize
%\resizebox{0.7\columnwidth}{0.08\textheight}{% % !
%\resizebox{0.7\columnwidth}{!}{% 
%\setlength\extrarowheight{2.4pt}
\begin{tabular}{|l|l|l|l|l|}
\hline  
\multicolumn{1}{|c|}{} 
&\multicolumn{1}{|c|}{A100} 
&\multicolumn{1}{|c|}{P100} 
&\multicolumn{1}{|c|}{MI250} 
&\multicolumn{1}{|c|}{{\funnyPbar}} \\ \cline{1-5}
\hline  
\hline  
\multicolumn{1}{|l|}{OpenACC} 
&\multicolumn{1}{|c|}{100\%} 
&\multicolumn{1}{|c|}{50\%} 
&\multicolumn{1}{|c|}{\st{66\%}, 50\%} 
&\multicolumn{1}{|c|}{\st{72\%}, 78\%} \\ \cline{1-5}
\hline  
\hline  
\multicolumn{1}{|l|}{OpenMP} 
&\multicolumn{1}{|c|}{75\%} 
&\multicolumn{1}{|c|}{100\%} 
&\multicolumn{1}{|c|}{\st{80\%}, 60 \%} 
&\multicolumn{1}{|c|}{\st{85\%}, 78\%} \\ \cline{1-5}
\hline  
\hline  
\multicolumn{1}{|l|}{Kokkos} 
&\multicolumn{1}{|c|}{50\%} 
&\multicolumn{1}{|c|}{33\%} 
&\multicolumn{1}{|c|}{\st{100\%}, 75\%} 
&\multicolumn{1}{|c|}{\st{61\%}, 53\%} \\ \cline{1-5}
\hline  
\hline  
\multicolumn{1}{|l|}{SYCL} 
&\multicolumn{1}{|c|}{60\%} 
&\multicolumn{1}{|c|}{62.5\%} 
&\multicolumn{1}{|c|}{100\%} 
&\multicolumn{1}{|c|}{74\%} \\ \cline{1-5}
\hline  
\end{tabular}
%}
%\end{center}
\label{table:t2}
\end{minipage}
\end{center}
\end{table*}

Table 1 (left) presents the runtimes, in seconds, of a fictional mini-app called CloverTree on two NVIDIA GPUs and one AMD GPU (A100, P100, and MI250) using three CloverTree implementations (OpenACC,
OpenMP, and Kokkos). 

Table 1 (right) presents the application efficiencies of the relevant platform-application pairs and the performance portability scores calculated by the {\funnyPbar} metric. For example, OpenACC achieves the best performance on the A100 GPU and therefore its application efficiency is 100\%, while the application efficiencies of OpenMP and Kokkos are calculated relative to the performance of OpenACC. The calculations of the application efficiencies on P100 and MI250 are done in the same way. Then, the {\funnyPbar} scores are calculated for each CloverTree implementation.

Now, let us add a SYCL implementation of CloverTree. Table 2 (left) presents the previous runtimes, in seconds, of CloverTree on three NVIDIA GPUs (A100, P100, and MI250) using three CloverTree implementations (OpenACC, OpenMP, and Kokkos), including the runtimes of the new SYCL implementation. Table 2 (right) presents the application efficiency of each platform-implementation pair and the performance portability scores of the four implementations calculated by the {\funnyPbar} metric. 

Now, let us pay close attention to the results obtained in Table 2 (right). The application efficiencies of OpenACC, OpenMP, and Kokkos on the A100 and P100 GPUs have not changed because the baseline performance on the A100 GPU remains that of OpenACC and the baseline performance on the P100 GPU remains that of OpenMP. Therefore, we needed only to calculate the application efficiency of SYCL on the A100 and P100 GPUs relative to the baseline performances.

However, in the case of the MI250 GPU there is a surprising result. Since the performance of SYCL on the MI250 GPU has better performance than OpenACC, OpenMP, and Kokkos, it becomes the baseline performance with an application efficiency of 100\% (instead of Kokkos). As a result, the application efficiencies of OpenACC, OpenMP, and Kokkos change as well as their performance portability scores. Table 2 (right) presents the new scores alongside the old scores.

Why are these results surprising and do not make sense? Because the runtimes of OpenACC, OpenMP, and Kokkos did not change at all and all we did was add to the table the performance of SYCL and nothing else. Not only does this not make sense, but it is also a violation of criterion 4 of the {\funnyPbar} metric (or any other performance portability metric because it is a fundamental criterion), which says that ``{\it the performance portability score increases or decreases if the performance increases or decreases on any platform in S.''} However, the performance of OpenACC, OpenMP, and Kokkos did not increase or decrease on any platform in $S$! This is clearly a contradiction.

The next section will present solutions to this problem.

\section{Solutions}

In this section we propose three practical solutions to solve the application efficiency problem introduced in the previous section.

\subsection {Solution No. 1}

Usually, it is a good practice to use both performance efficiencies, architectural and application, for calculating the performance portability of an application because each sheds light on performance portability from a different perspective. However, if the next two solutions for using the application efficiency are not applicable to your case, our recommendation is to use the architectural efficiency. The application efficiency problems that we described in the previous section never occur when using the architectural efficiency.

\begin{table*}%[!htb]
\begin{center}
\caption{CloverTree Mini-app Running Times (left) and \\Application Efficiencies \& Performance Portability (right) \\including SYCL, CUDA and HIP.}
%\vspace{.5cm}
\hspace{-0.7cm}
\setlength\extrarowheight{2.0pt}
\begin{minipage}{.25\linewidth}
%\caption{CloverTree Mini-app Running Times.}
%\begin{center}
%\footnotesize
%\resizebox{0.7\columnwidth}{0.08\textheight}{% % !
%\resizebox{0.7\columnwidth}{!}{% 
%\setlength\extrarowheight{2.5pt}
\begin{tabular}{|l|l|l|l|}
\hline  
\multicolumn{1}{|c|}{} 
&\multicolumn{1}{|c|}{A100} 
&\multicolumn{1}{|c|}{P100} 
&\multicolumn{1}{|c|}{MI250} \\ \cline{1-4}
\hline  
\hline  
\multicolumn{1}{|l|}{OpenACC} 
&\multicolumn{1}{|c|}{30} 
&\multicolumn{1}{|c|}{50} 
&\multicolumn{1}{|c|}{60} \\ \cline{1-4}
\hline  
\hline  
\multicolumn{1}{|l|}{OpenMP} 
&\multicolumn{1}{|c|}{40} 
&\multicolumn{1}{|c|}{25} 
&\multicolumn{1}{|c|}{50} \\ \cline{1-4}
\hline  
\hline  
\multicolumn{1}{|l|}{Kokkos} 
&\multicolumn{1}{|c|}{60} 
&\multicolumn{1}{|c|}{75} 
&\multicolumn{1}{|c|}{40} \\ \cline{1-4}
\hline  
\hline  
\multicolumn{1}{|l|}{SYCL} 
&\multicolumn{1}{|c|}{50} 
&\multicolumn{1}{|c|}{40} 
&\multicolumn{1}{|c|}{30} \\ \cline{1-4}
\hline  
\hline  
\multicolumn{4}{|c|}{Baseline Performance} \\ \cline{1-4}
\hline  
\hline  
\multicolumn{1}{|l|}{} 
&\multicolumn{1}{|c|}{10} 
&\multicolumn{1}{|c|}{10} 
&\multicolumn{1}{|c|}{10} \\ \cline{1-4}
\hline  
\multicolumn{1}{|l|}{} 
&\multicolumn{1}{|c|}{CUDA} 
&\multicolumn{1}{|c|}{CUDA} 
&\multicolumn{1}{|c|}{HIP} \\ \cline{1-4}
\hline 
\end{tabular}
%}
%\end{center}
\label{table:t1}
\end{minipage}
%\end{table}  
%\begin{table}
\hspace{1.5cm}
\begin{minipage}{.35\linewidth}
%\caption{CloverTree App. Efficiencies \& Performance Portability.}
%\begin{center}
%\footnotesize
%\resizebox{0.7\columnwidth}{0.08\textheight}{% % !
%\resizebox{0.7\columnwidth}{!}{% 
%\setlength\extrarowheight{2.1pt}
\begin{tabular}{|l|l|l|l|l|}
\hline  
\multicolumn{1}{|c|}{} 
&\multicolumn{1}{|c|}{A100} 
&\multicolumn{1}{|c|}{P100} 
&\multicolumn{1}{|c|}{MI250} 
&\multicolumn{1}{|c|}{{\funnyPbar}} \\ \cline{1-5}
\hline  
\hline  
\multicolumn{1}{|l|}{OpenACC} 
&\multicolumn{1}{|c|}{33\%} 
&\multicolumn{1}{|c|}{20\%} 
&\multicolumn{1}{|c|}{17\%} 
&\multicolumn{1}{|c|}{23\%} \\ \cline{1-5}
\hline  
\hline  
\multicolumn{1}{|l|}{OpenMP} 
&\multicolumn{1}{|c|}{25\%} 
&\multicolumn{1}{|c|}{40\%} 
&\multicolumn{1}{|c|}{20\%} 
&\multicolumn{1}{|c|}{28\%} \\ \cline{1-5}
\hline  
\hline  
\multicolumn{1}{|l|}{Kokkos} 
&\multicolumn{1}{|c|}{17\%} 
&\multicolumn{1}{|c|}{13\%} 
&\multicolumn{1}{|c|}{25\%} 
&\multicolumn{1}{|c|}{18\%} \\ \cline{1-5}
\hline  
\hline  
\multicolumn{1}{|l|}{SYCL} 
&\multicolumn{1}{|c|}{20\%} 
&\multicolumn{1}{|c|}{25\%} 
&\multicolumn{1}{|c|}{33\%} 
&\multicolumn{1}{|c|}{26\%} \\ \cline{1-5}
\hline  
\hline  
\multicolumn{5}{|c|}{Baseline Application Efficiency} \\ \cline{1-5}
\hline  
\hline  
\multicolumn{1}{|l|}{} 
&\multicolumn{1}{|c|}{100\%} 
&\multicolumn{1}{|c|}{100\%} 
&\multicolumn{1}{|c|}{100\%} 
&\multicolumn{1}{|c|}{} \\ \cline{1-5}
\hline  
\multicolumn{1}{|l|}{} 
&\multicolumn{1}{|c|}{CUDA} 
&\multicolumn{1}{|c|}{CUDA} 
&\multicolumn{1}{|c|}{HIP}
&\multicolumn{1}{|c|}{} \\ \cline{1-5}
\hline 
\end{tabular}
%}
%\end{center}
\label{table:t2}
\end{minipage}
\end{center}
\end{table*}  

\subsection{Solution No. 2}

It makes sense to calculate the performance portability of applications developed by performance portability frameworks like OpenMP, OpenACC, Kokkos, and Raja, which were designed a priori to provide performance portability to applications. On the other hand, in general, it does not make sense to calculate the performance portability of applications developed by low-level and non-portable parallel programming models such as CUDA. However, the performance portability of an application developed by a low-level, well-optimized, and non-portable parallel programming model such as CUDA can be used for reference. This is exactly the principal idea of our second proposal for solving the application efficiency problem. In our demonstration, we use the performance of the CUDA and HIP implementations of the application of interest as the baseline performance for calculating the application efficiency. 

The idea behind this solution stems from the assumption that the performances of implementations of applications which are developed by parallel programming models that are low-level and well-optimized, such as CUDA and HIP, outperform the performances of applications that are developed by high-level parallel programming models such as OpenACC and Kokkos. The practical meaning that stems from this basic assumption is that the application efficiency of low-level and well-optimized applications will always be 100\% even if, in the future, we add to the list of implementations of the performance portability frameworks an implementation of a new high-level parallel programming model or if the performance of the implementations improves over time.

Table 3 (left) presents the previous runtimes of OpenACC, OpenMP, Kokkos, and SYCL on the three platforms of interest together with the runtimes of the CUDA implementation. It can be observed that the CUDA implementation outperforms the other implementations on all three platforms by far. Therefore, in Table 3 (right) the application efficiency of CUDA is 100\% for the three platforms and it is assumed that it will remain as such for every CloverTree implementation developed by a high-level parallel programming model and for every new platform that will be added to Tables 3 (left) and 3 (right) in the future. Moreover, in the case that a new CloverTree implementation developed by a high-level performance portability framework will be added to Tables 3 (left) and 3 (right), the previous performance portability score of the performance portability framework already present in the tables will not change. In other words, criterion 4 will not be violated.

\subsection{Solution No. 3}

At the 2023 SBAC-PAD conference, Marowka proposed establishing an open repository of the performance portability of applications, benchmarks, and models \cite{Marowka5}. The motivation behind this proposition was to organize the research on performance portability in order to allow informed conclusions to be drawn in future studies. Furthermore, such an open repository will allow a rigid framework of rules and regulated measurement mechanisms to be maintained for future studies of performance portability, whose results will be stored in the open repository accessible to the HPC community.

One of the added values of such an open repository is that it essentially includes a solution to solve the application efficiency problem which satisfies the original definition.

For this purpose, we made a minor update to the original definition of the application efficiency as follows. 

\bigskip
\bigskip

\begin{quote}
{\bf Definition: application efficiency - modified}

\textit{ The achieved performance of a given portable application-platform pair, normalized relative to the best-known performance of any application's implementation on the same platform \textbf{ in the performance portability repository}.}
\end{quote}

The only addition we made to the original definition is the phrase ``{\it in the performance portability repository.''} Hence, we narrowed the search for the best-known performance of an application's implementation on the same platform, as the definition states, from the space of the entire universe to the space of the performance portability repository.

Over time, the performance portability repository will include a large number of implementations of applications in a wide variety of performance portability frameworks, including low-level, well-optimized, and non-portable parallel programming models on various types of platforms of different architectures as well as state-of-the-art compilers and backend compilers.

However, the big advantage of such a repository lies in the fact that it will be standardized, objective, and based on strict operating and reporting guidelines. Such guidelines will ensure a fair, comparable, and meaningful measure of performance portability, while the requirement for detailed disclosure of the obtained results and the configuration settings will ensure reproducibility of the reported results.

Moreover, since the repository is restricted to a rule-based and supervised framework, if an implementation with better performance enters the repository, the performance portability calculation of the relevant applications will be automatically updated. Such an automatic update is possible if dynamic web pages are used, such as those of a spreadsheet, which enable automatic updates of the calculation of a given function if one of its variables changes its value. Such a solution allows for a common performance reference in the repository at any point in time for all applications and benchmark suites. In this way, the database of performance portability reports will remain uniform and consistent while allowing an objective comparison between applications with the possibility of reproducing the various results.

\section{An Undesirable Solution}

In 2023, Rangel et al. \cite{Rangel} studied the performance portability of a cosmology application that was ported from CUDA and HIP to SYCL running on GPUs from three different vendors: NVIDIA (NVIDIA A100-SXM4-40GB), Intel (Intel Data Center GPU Max 1550), and AMD (AMD Instinct MI250X). The authors used the application efficiency approach and reported that their SYCL implementation achieved a performance portability of 96\%.
 
The optimization process of the application was carried out in stages and in a graduated manner. In the first step, a hotspot analysis was performed to identify kernels where the most time was spent during the application's execution. After that, several optimization techniques (variants) were applied to the kernels in order to obtain optimal performance. At the end of the process, it became clear to the researchers that they had run into a problem. Now, we present the problem as it was expressed by the authors in their own words and then we explain how it is related to the application efficiency calculation:

{\it``we cannot identify a single variant that delivers the best performance across all architectures and kernels highlights the difficulty of writing a single-source application that achieves high performance portability across diverse architectures. Even though all three architectures here are GPUs, running the same code, compiled with similar compilers, they still exhibit very different affinities for different variants of the same kernel.''}

In other words, the authors did not find a baseline performance that allowed them to calculate the application efficiency even with the method used among HPC researchers presented in Section 4. But we already know that they calculated and found that the performance portability of the application is 96\%, so what is the baseline performance they used?
 
To answer this intriguing question we return back to the text to present the solution found by the authors and in their own words:

{\it ``In all cases, application efficiency is calculated relative to a hypothetical application that is able to use the best version of each kernel on every platform, irrespective of source language or compiler.''}

At first reading, the idea of using a hypothetical application sounds intriguing and innovative. But on a second reading, it becomes clear that the authors did not present even one simple example of the hypothetical application they used. Therefore, we cannot assess how it is possible to derive from a given SYCL application its hypothetical version so that we are {\it ``able to use the best version of each kernel on every platform, regardless of source language or compiler.''} Furthermore, the lack of an example of such a hypothetical application makes it impossible to follow the calculations that the authors made, and it is impossible to reproduce their results, as is currently required from similar studies at scientific conferences, such as the conference where this study was presented. As long as we do not have full information about this solution, we cannot recommend using it.

\section{Portability Efficiency}

In this section we present a new performance efficiency approach called {\it portability efficiency}, which does not suffer from the biases of the application efficiency approach, as we demonstrated in previous sections. Furthermore, the {\it portability efficiency} approach, unlike other approaches, makes it possible to examine, based on the same data, the impact of each architecture on the performance portability of the entire application, and it better expresses the inherent connotation of {\it performance portability}. 

In the process of formulating the concept and definition of {\it portability efficiency}, we were inspired by the work of Sabne et al. \cite{Sabne1, Sabne2}, who studied the performance portability achieved in 12 OpenACC programs (four scientific kernels: Jacobi, Laplace, Matmul, and SPmul; alongside eight apps from the Rodinia benchmark suite \cite{Rodinia}: Srad, Hotspot, NW, LUD, BFS, Backprop, Kmeans, and CFD) on three different platforms (NVIDIA GTX 680, AMD Radeon HD 7970, and Intel Xeon Phi).

In order to enable a given OpenACC application to run on these three different architectures, the OpenARC compiler was developed \cite{OpenARC}. OpenARC is a source-to-source compiler for C-based OpenACC programs that translates an OpenACC program into a CUDA program for NVIDIA GPUs, and into OpenCL programs for AMD and Intel GPUs. It is based on intermediate representation, HeteroIR, for mapping high-level programming models to heterogeneous architectures. OpenARC contains a runtime system that performs compiler optimizations, such as loop unrolling and parallel loop swap, as well as an automatic tuning system that searches for the optimal program settings, such as the number of gangs and number of workers.

The study in \cite{Sabne1, Sabne2} examined the capabilities of the OpenARC compiler to translate a high-level programming model (e.g., OpenACC) to a low-level programming model (e.g., CUDA and OpenCL), while performing optimizations and automatic tunings for the different target architectures. From the obtained running results, it is also possible to deduce the level of performance portability that the OpenACC--OpenARC combination yields. 
On the other hand, we focus on finding better performance efficiency approaches for a more adequate calculation of performance portability of applications.
The new performance efficiency approach that we propose in this paper better expresses the performance efficiency of the application than the {\it application efficiency} approach and without the inherent flaws in its definition and in the way it is used as demonstrated in the previous sections.

The benchmarks show that the OpenACC--OpenARC combination achieves a performance portability of 76.5\% when the calculation is done using the {\funnyPbar} metric. These results are very encouraging.

The differences across accelerators and architectures may cause a well-set and optimized application aimed at achieving optimal performance on one architecture to run inefficiently on another architecture with the same application settings and optimizations. Therefore, the performance portability of an application is a function of the ratio of the achieved performance by the best-performing application settings on one architecture on the destination architecture. 

For example, let us consider the case that application $A$ moves from an NVIDIA GPU to an AMD GPU. After empirically finding the application settings that yield the best performance of application $A$ on the NVIDIA GPU, we run application $A$ with the same settings on the AMD GPU and measure a throughput of 90 GFLOPS. Next, we run application $A$ on the AMD GPU and find that the best achievable throughput is 100 GFLOPS. We will say in such a case that the performance efficiency  that application $A$ demonstrates is 90/100 = 90\%. We call this fraction {\it portability efficiency}.

\begin{table*}%[!htb]
\begin{center}
\caption{CloverTree performance (in GFLOPS) on destination platforms with best setting of the source platforms \\and the best performance on the destination platforms using OpenACC(left) and Kokkos(right)}
%\vspace{.5cm}
\setlength\extrarowheight{2.0pt}
\begin{minipage}{.25\linewidth}
%\caption{CloverTree Mini-app Running Times.}
%\begin{center}
%\footnotesize
%\resizebox{0.7\columnwidth}{0.08\textheight}{% % !
%\resizebox{0.7\columnwidth}{!}{% 
%\setlength\extrarowheight{2.5pt}
\begin{tabular}{|l|l|l|l|l|}
\hline  
\multicolumn{1}{|c|}{OpenACC} 
&\multicolumn{4}{|c|}{Destination Platform} \\ \cline{1-5}
\hline  
\hline  
\multicolumn{1}{|l|}{} 
&\multicolumn{1}{|c|}{} 
&\multicolumn{1}{|c|}{A100} 
&\multicolumn{1}{|c|}{P100} 
&\multicolumn{1}{|c|}{MI250} \\ \cline{1-5}
\hline  
\hline  
\multicolumn{1}{|c|}{Source} 
&\multicolumn{1}{|c|}{A100} 
&\multicolumn{1}{|c|}{--} 
&\multicolumn{1}{|c|}{60} 
&\multicolumn{1}{|c|}{36} \\ \cline{2-5}
%\hline  
%\hline  
\multicolumn{1}{|c|}{Platform} 
&\multicolumn{1}{|c|}{P100} 
&\multicolumn{1}{|c|}{88} 
&\multicolumn{1}{|c|}{--} 
&\multicolumn{1}{|c|}{42} \\ \cline{2-5}
%\hline  
%\hline  
\multicolumn{1}{|l|}{} 
&\multicolumn{1}{|c|}{MI250} 
&\multicolumn{1}{|c|}{80} 
&\multicolumn{1}{|c|}{60} 
&\multicolumn{1}{|c|}{--} \\ \cline{1-5}
\hline  
\hline  
\multicolumn{1}{|c|}{Best Perf.} 
&\multicolumn{1}{|c|}{} 
&\multicolumn{1}{|c|}{100} 
&\multicolumn{1}{|c|}{80} 
&\multicolumn{1}{|c|}{60} \\ \cline{1-5}
\hline 
\end{tabular}
%}
%\end{center}
\label{table:t1}
%\caption{Jacobi}
\end{minipage}
%\end{table}  
%\begin{table}
\hspace{3.0cm}
\begin{minipage}{.35\linewidth}
%\caption{CloverTree Mini-app Running Times.}
%\begin{center}
%\footnotesize
%\resizebox{0.7\columnwidth}{0.08\textheight}{% % !
%\resizebox{0.7\columnwidth}{!}{% 
%\setlength\extrarowheight{2.5pt}
\begin{tabular}{|l|l|l|l|l|}
\hline  
\multicolumn{1}{|c|}{Kokkos} 
&\multicolumn{4}{|c|}{Destination Platform} \\ \cline{1-5}
\hline  
\hline  
\multicolumn{1}{|l|}{} 
&\multicolumn{1}{|c|}{} 
&\multicolumn{1}{|c|}{A100} 
&\multicolumn{1}{|c|}{P100} 
&\multicolumn{1}{|c|}{MI250} \\ \cline{1-5}
\hline  
\hline  
\multicolumn{1}{|c|}{Source} 
&\multicolumn{1}{|c|}{A100} 
&\multicolumn{1}{|c|}{--} 
&\multicolumn{1}{|c|}{56} 
&\multicolumn{1}{|c|}{40} \\ \cline{2-5}
%\hline  
%\hline  
\multicolumn{1}{|c|}{Platform} 
&\multicolumn{1}{|c|}{P100} 
&\multicolumn{1}{|c|}{81} 
&\multicolumn{1}{|c|}{--} 
&\multicolumn{1}{|c|}{30} \\ \cline{2-5}
%\hline  
%\hline  
\multicolumn{1}{|l|}{} 
&\multicolumn{1}{|c|}{MI250} 
&\multicolumn{1}{|c|}{63} 
&\multicolumn{1}{|c|}{63} 
&\multicolumn{1}{|c|}{--} \\ \cline{1-5}
\hline  
\hline  
\multicolumn{1}{|c|}{Best Perf.} 
&\multicolumn{1}{|c|}{} 
&\multicolumn{1}{|c|}{90} 
&\multicolumn{1}{|c|}{70} 
&\multicolumn{1}{|c|}{50} \\ \cline{1-5}
\hline 
\end{tabular}
%}
%\end{center}
\label{table:t1}
%\caption{Jacobi}
\end{minipage}
\end{center}
\end{table*}  

%=====================================

\begin{table*}%[!htb]
\begin{center}
\caption{CloverTree portability efficiencies and performance portability \\ implemented using OpenACC(left) and Kokkos(right)}
%\vspace{.5cm}
\setlength\extrarowheight{2.0pt}
\begin{minipage}{.25\linewidth}
%\caption{CloverTree Mini-app Running Times.}
%\begin{center}
%\footnotesize
%\resizebox{0.7\columnwidth}{0.08\textheight}{% % !
%\resizebox{0.7\columnwidth}{!}{% 
%\setlength\extrarowheight{2.5pt}
\begin{tabular}{|l|l|l|l|l|}
\hline  
\multicolumn{1}{|c|}{OpenACC} 
&\multicolumn{4}{|c|}{Destination Platform} \\ \cline{1-5}
\hline  
\hline  
\multicolumn{1}{|l|}{} 
&\multicolumn{1}{|c|}{} 
&\multicolumn{1}{|c|}{A100} 
&\multicolumn{1}{|c|}{P100} 
&\multicolumn{1}{|c|}{MI250} \\ \cline{1-5}
\hline  
\hline  
\multicolumn{1}{|c|}{Source} 
&\multicolumn{1}{|c|}{A100} 
&\multicolumn{1}{|c|}{--} 
&\multicolumn{1}{|c|}{75\%} 
&\multicolumn{1}{|c|}{60\%} \\ \cline{2-5}
%\hline  
%\hline  
\multicolumn{1}{|c|}{Platform} 
&\multicolumn{1}{|c|}{P100} 
&\multicolumn{1}{|c|}{88\%} 
&\multicolumn{1}{|c|}{--} 
&\multicolumn{1}{|c|}{70\%} \\ \cline{2-5}
%\hline  
%\hline  
\multicolumn{1}{|l|}{} 
&\multicolumn{1}{|c|}{MI250} 
&\multicolumn{1}{|c|}{80\%} 
&\multicolumn{1}{|c|}{75\%} 
&\multicolumn{1}{|c|}{--} \\ \cline{1-5}
\hline  
\hline  
\multicolumn{1}{|c|}{{\funnyPbar} Arch.} 
&\multicolumn{1}{|c|}{} 
&\multicolumn{1}{|c|}{84\%} 
&\multicolumn{1}{|c|}{75\%} 
&\multicolumn{1}{|c|}{65\%} \\ \cline{1-5}
\hline  
\multicolumn{1}{|c|}{{\funnyPbar}} 
&\multicolumn{4}{|c|}{74.66\%} \\ \cline{1-5}
\hline 
\end{tabular}
%}
%\end{center}
\label{table:t1}
%\caption{Jacobi}
\end{minipage}
%\end{table}  
%\begin{table}
\hspace{3.0cm}
\begin{minipage}{.35\linewidth}
%\caption{CloverTree Mini-app Running Times.}
%\begin{center}
%\footnotesize
%\resizebox{0.7\columnwidth}{0.08\textheight}{% % !
%\resizebox{0.7\columnwidth}{!}{% 
%\setlength\extrarowheight{2.5pt}
\begin{tabular}{|l|l|l|l|l|}
\hline  
\multicolumn{1}{|c|}{Kokkos} 
&\multicolumn{4}{|c|}{Destination Platform} \\ \cline{1-5}
\hline  
\hline  
\multicolumn{1}{|l|}{} 
&\multicolumn{1}{|c|}{} 
&\multicolumn{1}{|c|}{A100} 
&\multicolumn{1}{|c|}{P100} 
&\multicolumn{1}{|c|}{MI250} \\ \cline{1-5}
\hline  
\hline  
\multicolumn{1}{|c|}{Source} 
&\multicolumn{1}{|c|}{A100} 
&\multicolumn{1}{|c|}{--} 
&\multicolumn{1}{|c|}{80\%} 
&\multicolumn{1}{|c|}{80\%} \\ \cline{2-5}
%\hline  
%\hline  
\multicolumn{1}{|c|}{Platform} 
&\multicolumn{1}{|c|}{P100} 
&\multicolumn{1}{|c|}{90\%} 
&\multicolumn{1}{|c|}{--} 
&\multicolumn{1}{|c|}{60\%} \\ \cline{2-5}% mmmmm
%\hline  
%\hline  
\multicolumn{1}{|l|}{} 
&\multicolumn{1}{|c|}{MI250} 
&\multicolumn{1}{|c|}{70\%} 
&\multicolumn{1}{|c|}{90\%} 
&\multicolumn{1}{|c|}{--} \\ \cline{1-5}
\hline  
\hline  
\multicolumn{1}{|c|}{{\funnyPbar} Arch.} 
&\multicolumn{1}{|c|}{} 
&\multicolumn{1}{|c|}{80\%} 
&\multicolumn{1}{|c|}{85\%} 
&\multicolumn{1}{|c|}{70\%} \\ \cline{1-5}
\hline  
\multicolumn{1}{|c|}{{\funnyPbar}} 
&\multicolumn{4}{|c|}{78.33\%} \\ \cline{1-5}
\hline 
\end{tabular}
%}
%\end{center}
\label{table:t1}
%\caption{Jacobi}
\end{minipage}
\end{center}
\end{table*}  
%\newpage

The portability efficiency approach explicitly defines the performance cost involved in a one-way migration of the application from a source platform to a target platform and thus succeeds in better expressing the portability aspect in the overall calculation of the performance portability better than any other performance efficiency approach proposed in the literature to the best of our knowledge.

Formally, for a given base platform $b$ and a target platform $t$, the portability efficiency $\eta$ of application $a$ solving problem $p$ on platform $t$, when it is transferred from platform $b$, is defined as the achieved performance of application a on platform $t$, with the best-performing application setting on platform $b$, $C(b)$, as fraction of the best observed performance of application $a$ on platform $t$:

\begin{align}
{\eta(a, p, b \rightarrow t) = \frac{P_{d,C(a)}} {P_{t,C(t)}} }
\end{align}

Now, let us demonstrate how to calculate the performance portability using {\funnyPbar} metric and portability efficiency approach.

First, for a given supported set of platforms  $S \subseteq H$ where $|S| >1$, the cardinality of a set $D$ of ordered selections (permutations) of two platforms from a set of $|S|$ platforms is:

\begin{align}
{D = 2 \cdot  {|S|\choose 2} = \frac{|S|!} {(|S| - 2)!)} }
\end{align}

Hence, the performance portability of a given application $a$ solving problem $p$ using portability efficiency $\eta$ is: 

\begin{align}
{\funnyPbar}(a, p, S, H) = 
\begin{cases}
  \frac{\sum_{\{i,j\} \in D}^{} {[\eta (a, p, i \rightarrow j)]}} {|D|}     & \text{if $|S| > 1$} \\ 
  \text{$0$}                                      & \text{otherwise} 
\end{cases}
\end{align}

\smallskip
Let us look on the following hypothetical example.
Table 4 presents the performance, in GFLOPS, of our fictional CloverTree mini-app implementations using OpenACC (left) and Kokkos (right) on two NVIDIA GPUs (A100 and P100) and on AMD GPU (MI250).
For example, the performance of the OpenACC implementation on A100 GPU with the best setting of P100 GPU is 88 GFLOPS and with the best setting of MI250 GPU is 80 GFLOPS while its best performance on A100 GPU is 100 GFLOPS.

Table 5 (left) presents the {\it portability efficiency} scores, based on the performance results of Table 4 (left), of the OpenACC implementation of the CloverTree application. For example, the {\it portability efficiency} score of the CloverTree application when it is transferred from A100 to MI250 is 60\%. Conversely, when it is transferred from MI250 to A100 the {\it portability efficiency} score achieved is 80\%.    

The total {\it performance portability} score, 74.66\%, is calculated by the {\funnyPbar} metric, which is actually the arithmetic mean of the {\it portability efficiency} scores of all the source--destination platform pairs in $H$.  

Now, let us add the results of Kokkos implementation to Tables 4 and 5.
Table 5 (right) presents the portability efficiency scores, based on the performance results of Table 4 (right), of the Kokkos implementation of the CloverTree application.

\begin{table*}%[!htb]
\begin{center}
\caption{Jacobi (left) and CFD (right)  Portability Efficiencies and Performance Portability \\ on NVIDIA, AMD and Intel accelerators implemented using OpenACC-OpenARC .}
%\vspace{.5cm}
\setlength\extrarowheight{2.0pt}
\begin{minipage}{.25\linewidth}
%\caption{CloverTree Mini-app Running Times.}
%\begin{center}
%\footnotesize
%\resizebox{0.7\columnwidth}{0.08\textheight}{% % !
%\resizebox{0.7\columnwidth}{!}{% 
%\setlength\extrarowheight{2.5pt}
\begin{tabular}{|l|l|l|l|l|}
\hline  
\multicolumn{1}{|c|}{Jacobi} 
&\multicolumn{4}{|c|}{Destination Platform} \\ \cline{1-5}
\hline  
\hline  
\multicolumn{1}{|l|}{} 
&\multicolumn{1}{|c|}{} 
&\multicolumn{1}{|c|}{NVIDIA} 
&\multicolumn{1}{|c|}{AMD} 
&\multicolumn{1}{|c|}{Intel} \\ \cline{1-5}
\hline  
\hline  
\multicolumn{1}{|c|}{Source} 
&\multicolumn{1}{|c|}{NVIDIA} 
&\multicolumn{1}{|c|}{--} 
&\multicolumn{1}{|c|}{93\%} 
&\multicolumn{1}{|c|}{65\%} \\ \cline{2-5}
%\hline  
%\hline  
\multicolumn{1}{|c|}{Platform} 
&\multicolumn{1}{|c|}{AMD} 
&\multicolumn{1}{|c|}{95\%} 
&\multicolumn{1}{|c|}{--} 
&\multicolumn{1}{|c|}{85\%} \\ \cline{2-5}
%\hline  
%\hline  
\multicolumn{1}{|l|}{} 
&\multicolumn{1}{|c|}{Intel} 
&\multicolumn{1}{|c|}{91\%} 
&\multicolumn{1}{|c|}{83\%} 
&\multicolumn{1}{|c|}{--} \\ \cline{1-5}
\hline  
\hline  
\multicolumn{1}{|c|}{{\funnyPbar} Arch.} 
&\multicolumn{1}{|c|}{} 
&\multicolumn{1}{|c|}{93\%} 
&\multicolumn{1}{|c|}{88\%} 
&\multicolumn{1}{|c|}{75\%} \\ \cline{1-5}
\hline  
\multicolumn{1}{|c|}{{\funnyPbar}} 
&\multicolumn{4}{|c|}{85.33\%} \\ \cline{1-5}
\hline 
\end{tabular}
%}
%\end{center}
\label{table:t1}
%\caption{Jacobi}
\end{minipage}
%\end{table}  
%\begin{table}
\hspace{3.0cm}
\begin{minipage}{.35\linewidth}
%\begin{center}
%\footnotesize
%\resizebox{0.7\columnwidth}{0.08\textheight}{% % !
%\resizebox{0.7\columnwidth}{!}{% 
%\setlength\extrarowheight{2.1pt}
\begin{tabular}{|l|l|l|l|l|}
\hline  
\multicolumn{1}{|c|}{CFD} 
&\multicolumn{4}{|c|}{Destination Platform} \\ \cline{1-5}
\hline  
\hline  
\multicolumn{1}{|l|}{} 
&\multicolumn{1}{|c|}{} 
&\multicolumn{1}{|c|}{NVIDIA} 
&\multicolumn{1}{|c|}{AMD} 
&\multicolumn{1}{|c|}{Intel} \\ \cline{1-5}
\hline  
\hline  
\multicolumn{1}{|c|}{Source} 
&\multicolumn{1}{|c|}{NVIDIA} 
&\multicolumn{1}{|c|}{--} 
&\multicolumn{1}{|c|}{93\%} 
&\multicolumn{1}{|c|}{31\%} \\ \cline{2-5}
%\hline  
%\hline   
\multicolumn{1}{|c|}{Platform} 
&\multicolumn{1}{|c|}{AMD} 
&\multicolumn{1}{|c|}{63\%} 
&\multicolumn{1}{|c|}{--} 
&\multicolumn{1}{|c|}{61\%} \\ \cline{2-5}
%\hline  
%\hline  
\multicolumn{1}{|l|}{} 
&\multicolumn{1}{|c|}{Intel} 
&\multicolumn{1}{|c|}{15\%} 
&\multicolumn{1}{|c|}{15\%} 
&\multicolumn{1}{|c|}{--} \\ \cline{1-5}
\hline  
\hline  
\multicolumn{1}{|c|}{{\funnyPbar} Arch.} 
&\multicolumn{1}{|c|}{} 
&\multicolumn{1}{|c|}{39\%} 
&\multicolumn{1}{|c|}{54\%} 
&\multicolumn{1}{|c|}{46\%} \\ \cline{1-5}
\hline  
\multicolumn{1}{|c|}{{\funnyPbar}} 
&\multicolumn{4}{|c|}{46.33\%} \\ \cline{1-5}
\hline 
\end{tabular}
%}
%\end{center}
\label{table:t2}
%\caption{CFD}
\end{minipage}
\end{center}
\end{table*}  

Now, let us pay close attention to the fact that by adding the performance results of Kokkos implementation in Table 4 (right) and the calculations of the portability efficiencies and performance portability scores in table 5 (right), the results of the performance portability scores of OpenACC implementation have not affected in any way contrary to the calculation method of the application efficiency approach which caused the performance portability scores to be updated  as the examples in Section 4 demonstrate.

The reason for this is that the measurement and calculation of the portability efficiency of a given application does not depend on the performance of another application but only on its own performance for different application settings on the set of platforms of interest.

Table 5 also provides interesting diagnoses of {\it performance portability} from the perspective of the different architectures. For example, it can be seen, in Table 5 (left), that it is advisable to transfer the application from A100 to P100 (75\%) rather than to MI250 (60\%) because a better {\it portability efficiency} score is achieved. Another insight that emerges from the data of Table 5 (left) is that, on average, A100 (84\%) contributes more to the performance portability of CloverTree rather than P100 (75\%) or MI250 (65\%). Such information is especially important in critical heterogeneous systems, which are required to work continuously without interruption and to react in real time in cases where is necessary to perform an ad hoc migration of an application from one platform to another as a result of a failure or some other reason.

Next, we present the results of two real applications. Table 6 shows the portability efficiency and the performance portability scores of the OpenACC Jacobi application (left) and the OpenACC CFD application (right). First, it can be observed that the Jacobi application achieves a performance portability score (85.33\%) that is two times better than achieved with the CFD application (46.33\%) on the tested platforms (NVIDIA GTX 680, AMD Radeon HD 7970, and Intel Xeon Phi). In the case of the Jacobi application, it can be seen that it is advisable to transfer the application from NVIDIA to AMD (93\%) or from AMD to NVIDIA (95\%), in order to preserve maximum performance portability. Moreover, these platforms contribute more to the total performance portability of Jacobi while demonstrating performance portability scores of 93\% and 88\% compared to Intel's 75\%. Similarly, in the case of the CFD application, the transition from the NVIDIA to AMD platforms, and in the opposite direction, is the most advisable.  On the other hand, these platforms do not show a particularly noticeable contribution to the total performance portability of the CFD application.

\section{Conclusions}

Application efficiency is an attractive approach to calculating the performance portability of an application because it is simple and easy to use.
In this paper we demonstrated that the method of using this approach of performance efficiency yields calculations that do not align with the expectations inherent in its formal definition, and it violates the criteria of current performance portability metrics.

Fortunately, there are solutions that make the use of application efficiency possible without side effects and which satisfy the formal definition without violating the performance portability criteria. We proposed three practical solutions, two of which are local solutions. The third solution is a global solution with many additional advantages.

Finally, we proposed a new performance efficiency approach called {\it portability efficiency}. This approach is immune to the application efficiency problem, better expresses the aspect of portability, and it allows the user to explore different aspects of the impact of the different architectures on the performance portability of the application.

We hope that the solutions we proposed will help the HPC community research to enhance studies in the field of performance portability.

\end{document}